\documentclass[conference, a4paper]{IEEEtran}
\IEEEoverridecommandlockouts
\usepackage{cite}
\usepackage{amsmath,amssymb,amsfonts,amsthm}
\usepackage{mathtools}
\usepackage{graphicx}
\usepackage{textcomp}
\usepackage{xcolor}
\usepackage{bm}
\usepackage{multirow}
\usepackage{subfigure}

\allowdisplaybreaks

\makeatletter
 \def\Hline{%
 \noalign{\ifnum0=`}\fi\hrule \@height 1.2pt \futurelet
 \reserved@a\@xhline}
\makeatother

\newtheorem{thm}{Theorem}

\newcommand{\tsum}{\textstyle\sum}
\def\BibTeX{{\rm B\kern-.05em{\sc i\kern-.025em b}\kern-.08em
    T\kern-.1667em\lower.7ex\hbox{E}\kern-.125emX}}
\begin{document}

\title{Audio Spotforming Using Nonnegative Tensor Factorization with Attractor-Based Regularization}

\author{\IEEEauthorblockN{Shoma Ayano\IEEEauthorrefmark{1},
Li Li\IEEEauthorrefmark{2}, Shogo Seki\IEEEauthorrefmark{2}, and
Daichi Kitamura\IEEEauthorrefmark{1}}
\IEEEauthorblockA{\IEEEauthorrefmark{1}National Institute of Technology, Kagawa, Japan\\
\IEEEauthorrefmark{2}CyberAgent, Inc., Japan}}

\maketitle

\begin{abstract}
Spotforming is a target-speaker extraction technique that uses multiple microphone arrays.
This method applies beamforming (BF) to each microphone array, and the common components among the BF outputs are estimated as the target source.
This study proposes a new common component extraction method based on nonnegative tensor factorization (NTF) for higher model interpretability and more robust spotforming against hyperparameters.
Moreover, attractor-based regularization was introduced to facilitate the automatic selection of optimal target bases in the NTF.
Experimental results show that the proposed method performs better than conventional methods in spotforming performance and also shows some characteristics suitable for practical use.
\end{abstract}


\section{Introduction}

Target speaker extraction extracts only the target source from the observed signal. 
This technique can be applied to front-end systems in various audio applications, including automatic speech recognition.

\begin{figure}[t]
    \begin{center}
        \includegraphics[width=0.98\columnwidth]{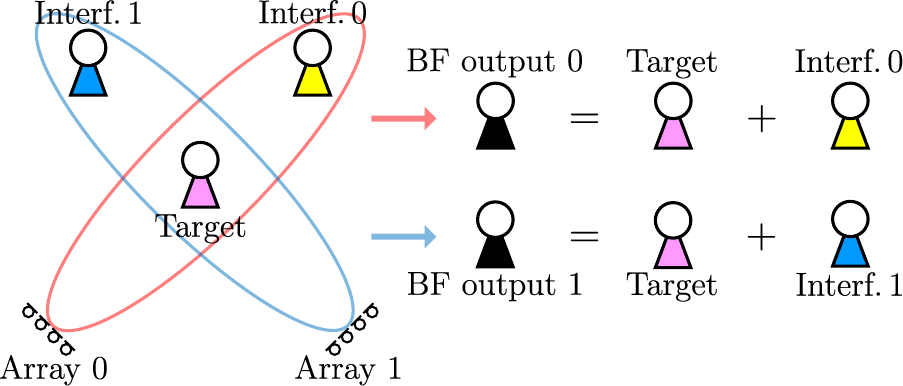}
        \vspace{-10pt}
    \end{center}
    \caption{Situations and signals estimated by two BF filters.}
    \label{fig:Model}
    \vspace{-5pt}
\end{figure}

Beamforming (BF)~\cite{Brandstein2001_micArrays} is the most common approach for target speaker extraction when using a microphone array.
Because BF emphasizes all source signals present in a specific direction from the microphone array, interference sources in the same direction cannot be suppressed. 
\textit{Spotforming}~\cite{Taseska2016_spotforming} using multiple microphone arrays was proposed to solve this problem. 
Spotforming aims to extract only the target source from a specific area, as shown in Fig.~\ref{fig:Model}.
A spatial spotforming filter using all synchronized microphone arrays was proposed~\cite{Taseska2016_spotforming}.
In addition, the optimal arrangement of multiple microphone arrays for spotforming was investigated~\cite{Sekiguchi2017_spotforming}.

\begin{figure}[t]
    \begin{center}
    \includegraphics[width=0.98\columnwidth]{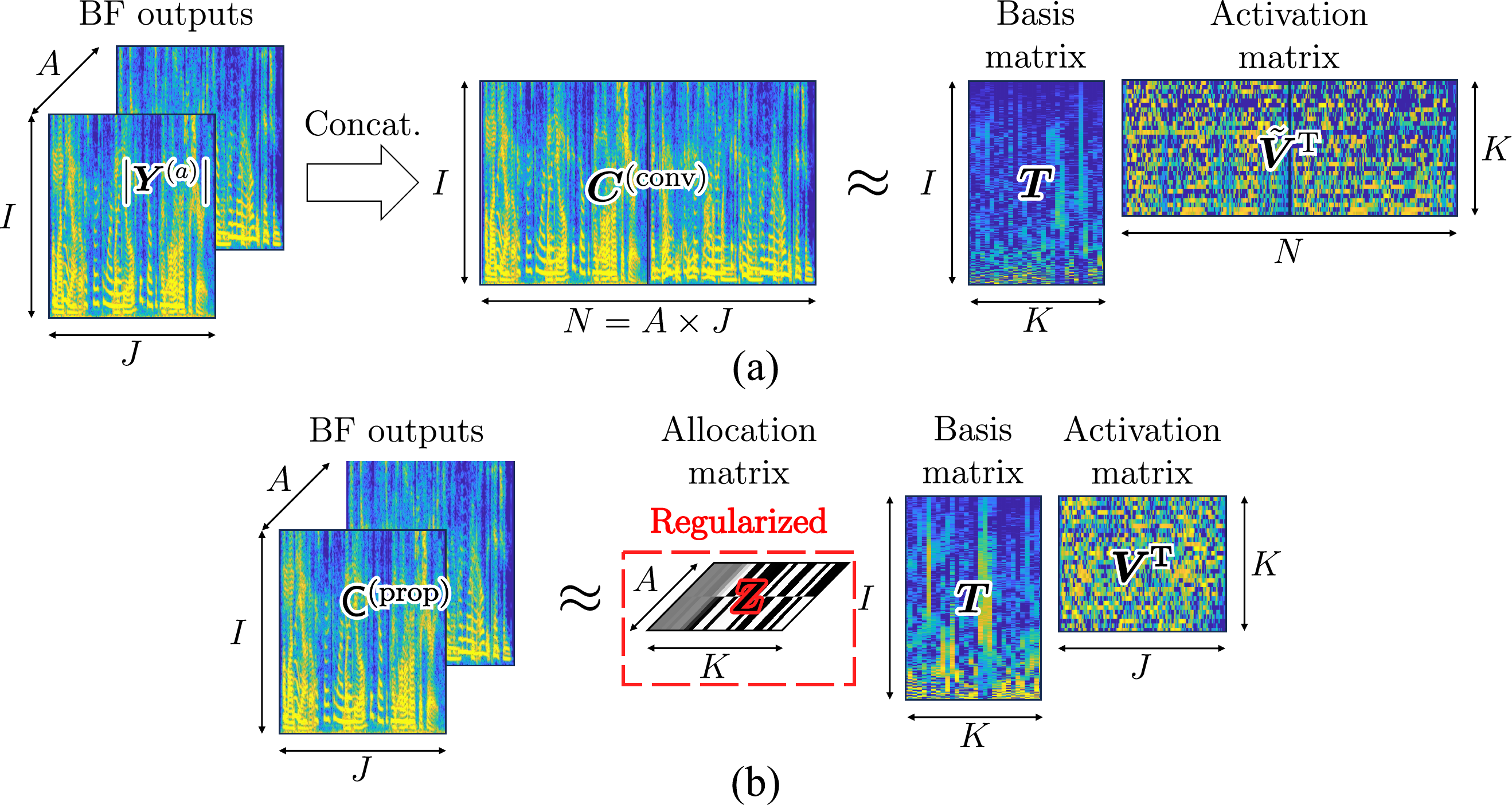}
    \end{center}
    \vspace{-10pt}
    \caption{Decomposition models of (a) NMF in the conventional method and (b) NTF in the proposed method.}
    \label{fig:Method}
    \vspace{-5pt}
\end{figure}

As another effective approach, spotforming utilizing nonnegative matrix factorization (NMF)~\cite{Lee1999_nmf} has been proposed~\cite{Kagimoto2022_spotforming}.
This method emphasizes the target directions using each microphone array with a BF. Then, it applies NMF~\cite{Lee1999_nmf} to the BF outputs concatenated in the time-frame dimension, as shown in Fig.~\ref{fig:Method}~(a). 
The target source is estimated as the common component extracted from the NMF decomposition results using a binary mask obtained by thresholding the activation matrix.
However, this decomposition lacks model interpretability because of the absence of explicit modeling of the relationship between each basis vector and each of the BF outputs, making it difficult to introduce effective regularization to enhance discriminative basis learning.
Moreover, the performance depends on the setting of the hyperparameters, including the number of basis vectors and the threshold value for extraction of common components.
These hyperparameters must be tuned in advance based on the observed signal characteristics, such as the signal length and time-frequency structures of the target and interference sources.

In this study, we propose a new spotforming method that has higher model interpretability and achieves more robust spotforming against hyperparameters. 
To this end achieve, we utilize nonnegative tensor factorization (NTF)~\cite{Cichocki2007_ntf} for common component extraction, as shown in Fig.~\ref{fig:Method}~(b). 
An allocation matrix is introduced to represent the basis vectors corresponding to the target source components, resulting in higher model interpretability. 
In addition, this approach enables the incorporation of attractor-based regularization to simultaneously facilitate discriminative basis learning and automatic optimization of the number of basis vectors for each source.
While our proposed method involves two hyperparameters, the number of basis vectors and the weight coefficient of the regularizer, the spotforming performance is shown to be relatively robust to variations in these values.

\section{Spotforming Using Multiple Microphone Arrays}

\subsection{Scenario of Spotforming and Its Signal Model}

In this study, we consider the scenario depicted in Fig.~\ref{fig:Model}. 
The target and interference sources are located in the same direction relative to each microphone array. 
Although the BF in each microphone array can enhance the target source, the interference sources in the same direction are also enhanced. 
However, because the residual interference sources are different for each BF output, estimation of the target source becomes feasible by extracting only the common components among the BF output signals.

Let $\mathsf{X}^{(a)}\in\mathbb{C}^{I\times J\times M}$ denote the time-frequency components of the multichannel observed signal captured by $a$-th microphone array, with its elements defined as $x_{i,j,m}^{(a)}$, where $i=0, 1, \cdots, I-1$, $j=0, 1, \cdots, J-1$, $a=0,1,\cdots,A-1$, and $m=0,1,\cdots,M-1$ are the indices of frequency bins, time frames, microphone arrays, and microphones within each array, respectively.
In spotforming using distributed microphone arrays, BF is initially applied within each microphone array as $\bm{Y}^{(a)} = f_{\theta_a}(\mathsf{X}^{(a)})$, where $\bm{Y}^{(a)}\in\mathbb{C}^{I\times J}$ represents the output spectrogram of the BF operation and $\theta_a$ denotes a specific look direction.
We also define the tensor for $(\bm{Y}^{(a)})_{a=0}^{A-1}$ as $\mathsf{Y}\in\mathbb{C}^{I\times J\times A}$, with its elements denoted as $y_{i,j}^{(a)}$.
The objective of spotforming is to estimate the components of the target source from $\mathsf{Y}$ that correspond to the common components of the BF outputs $(\bm{Y}^{(a)})_{a=0}^{A-1}$.

\subsection{Conventional NMF-Based Spotforming}

In the conventional method~\cite{Kagimoto2022_spotforming}, NMF extracts common components.
Let $\bm{C}^{(\mathrm{conv})}\in\mathbb{R}_{\geq 0}^{I\times N}$ denote an input matrix of NMF, with its elements defined as $c_{i, n}^{(\mathrm{conv})}$, where $n=0, 1, \cdots, N-1$ is the index of columns. 
This matrix is constructed by concatenating the amplitude spectrograms $(|\bm{Y}^{(a)}|)_{a=0}^{A-1}$ in the time-frame dimension, as shown in Fig.~\ref{fig:Method}~(a), namely, 
\begin{align}
    c^\mathrm{(conv)}_{i,n} \coloneqq c^\mathrm{(conv)}_{i,aJ+j} = \left|y^{(a)}_{i,j}\right|, \label{eq:convMatrix}
\end{align}
where $|\cdot|$ for matrices denotes an element-wise absolute operation and $N=AJ$.
Matrix $\bm{C}^{(\mathrm{conv})}$ is decomposed using NMF as follows:
\begin{align}
    \bm{C}^\mathrm{(conv)} \approx \bm{T}\tilde{\bm{V}}^\mathrm{T}~~~(c_{i,n}^\mathrm{(conv)} \approx \tsum_k t_{i,k} \tilde{v}_{n, k}), \label{eq:convNmf}
\end{align}
where $\bm{T} \in \{\bm{T} \in [0, 1]^{I\times K} | \sum_i t_{i,k}=1 \}$ and $\tilde{\bm{V}} \in \mathbb{R}_{\geq 0}^{N\times K}$ are the basis and activation matrices, respectively, and $k=0, 1,\cdots, K-1$ is the index of NMF basis vectors.
The optimization problem of \eqref{eq:convNmf} is defined as follows:
\begin{align}
     \underset{\bm{T}, \tilde{\bm{V}}}{\mathrm{minimize}}~&\tsum_{i,n}\mathcal{D}\left( c^\mathrm{(conv)}_{i,n} | \tsum_k t_{i,k}\tilde{v}_{n,k} \right) \nonumber \\
     \mathrm{s.t.}~&t_{i,k}, \tilde{v}_{n,k}\geq0~\forall i,k,n, 
    \label{eq:convCost}
\end{align}
where $\mathcal{D}$ is a divergence function.

After the estimation of $\bm{T}$ and $\tilde{\bm{V}}$, a binary mask matrix $\tilde{\bm{H}}\in\{0, 1\}^{J\times K}$ is calculated as
\begin{align}
    \tilde{h}_{j,k} =
    \begin{cases}
        1 & ~(\mathrm{if}~\tilde{v}_{aJ + j, k} > \tau ~\forall a) \\
        0 & ~(\mathrm{otherwise})
    \end{cases}, \label{eq:convMask}
\end{align}
where $\tilde{h}_{j,k}$ is an element of $\tilde{\bm{H}}$ and $\tau \geq 0$ is a thresholding value.
This binary mask defines the component with an activation greater than $\tau$ in all the microphone arrays as the common target source component and sets it to unity.

The spectrogram of the target source $\hat{\bm{S}}^{(a)}\in\mathbb{C}^{I\times J}$ can be estimated by a Wiener filter using the binary mask $\tilde{\bm{H}}$ as\footnote{The method proposed in~\cite{Kagimoto2022_spotforming} does not apply~\eqref{eq:convWiener}, but directly uses the estimated NMF model for obtaining $\hat{\mathbf{s}}^{(a)}$ with a phase recovery technique. Since~\eqref{eq:convWiener} slightly improves the performance, we employ~\eqref{eq:convWiener} in this paper.}
\begin{align}
    \hat{s}^{(a)}_{i,j} = \frac{\tsum_k{(t_{i,k}\tilde{h}_{j,k}\tilde{v}_{aJ+j,k}})^2}{\tsum_k{(t_{i,k}\tilde{v}_{aJ+j,k}})^2}y^{(a)}_{i,j}, \label{eq:convWiener}
\end{align}
where $\hat{s}^{(a)}_{i,j}$ is the $(i, j)$-th element of $\hat{\bm{S}}^{(a)}$.
To obtain the time domain signals $\hat{\mathbf{s}}^{(a)}\in\mathbb{R}^L$, an inverse short-time Fourier transform (STFT) is applied to $\hat{\bm{S}}^{(a)}$, where $L$ is the signal length of $\hat{\mathbf{s}}^{(a)}$.
Finally, a delay-and-sum operation using $(\hat{\mathbf{s}}^{(a)})_{a=0}^{A-1}$ is performed for further enhancement of the target source signal.

\section{Proposed Method}

\subsection{Motivations} \label{sect:motivation}
The NMF model~in Fig.~\ref{fig:Method}~(a) lacks model interpretability because of the absence of explicit modeling of the relationship between each basis vector in $\bm{T}$ and each of the BF outputs $(\bm{Y}^{(a)})_{a=0}^{A-1}$.
Consequently, it is difficult to regularize $\bm{T}$ or $\bm{V}$ to enhance discrimination between the target and interference source components. 
Moreover, the hyperparameters $K$ and $\tau$ must be appropriately tuned depending on the observed signal in advance, which increases the difficulty of putting the conventional method into practical use.

To address these shortcomings, we propose a novel spotforming method that utilizes an NTF to model the spectrograms of the BF outputs (Fig.~\ref{fig:Method}~(b)).
The proposed method allocates $K$ basis vectors in $\bm{T}$ to each microphone array using an allocation matrix $\bm{Z}$.
Furthermore, we introduce an attractor-based regularization into $\bm{Z}$ to allocate each basis vector automatically assign/allocate to the corresponding BF outputs. 
This significantly increases the interpretability of the model and makes basis vectors more discriminative than the conventional method.
This regularization automatically optimizes the number of basis vectors for the target source, resulting in robust spotforming against hyperparameter settings.

\subsection{NTF-Based Spotforming}

In the proposed method, the input tensor $\mathsf{C}^{(\mathrm{prop})}\in\mathbb{R}_{\geq 0}^{A\times I\times J}$ of the NTF is defined as follows (Fig.~\ref{fig:Method}~(b)):
\begin{align}
    c^\mathrm{(prop)}_{a,i,j} \coloneqq \left|y^{(a)}_{i,j}\right|, \label{eq:propTensor}
\end{align}
where $c^\mathrm{(prop)}_{a,i,j}$ is an element of $\mathsf{C}^{(\mathrm{prop})}$.
In contrast to~\eqref{eq:convMatrix}, the three-dimensional tensor $\mathsf{C}^{(\mathrm{prop})}$ maintains the physical dimensions of the microphone array, frequency bins, and time frames.
Then, $\mathsf{C}^{(\mathrm{prop})}$ is decomposed into three matrices, i.e., the allocation matrix $\bm{Z} = [\bm{z}_0, \bm{z}_1, \cdots, \bm{z}_{K-1}]\in\{\bm{Z} \in [0, 1]^{A\times K} | \sum_a z_{a,k}=1 \}$, the basis matrix $\bm{T}=[\bm{t}_0, \bm{t}_1, \cdots, \bm{t}_{K-1}]$, and the activation matrix $\bm{V}\in\mathbb{R}_{\geq 0}^{J\times K}$, as
\begin{align}
    c_{a,i,j}^{(\mathrm{prop})} \approx \tsum_k z_{a,k}t_{i,k}v_{j,k}, \label{eq:propNtf}
\end{align}
where $z_{a, k}$ and $v_{j, k}$ are the elements of $\bm{Z}$ and $\bm{V}$, respectively, and $\bm{z}_k$ and $\bm{t}_k$ are the column vectors of $\bm{Z}$ and $\bm{T}$, respectively.

Matrix $\bm{Z}$ allocates $K$ basis vectors to $A$ microphone arrays to approximate input tensor $\mathsf{C}^{(\mathrm{prop})}$.
Because the target source components are commonly included across all BF outputs $(|\bm{Y}^{(a)}|)_{a=0}^{A-1}$, such basis vectors should be allocated to all microphone arrays. 
Therefore, if $\bm{t}_k$ represents the target source, $\bm{z}_k$ should be $\bm{z}_k \approx [1/A, 1/A, \cdots, 1/A]^\mathrm{T}$.
In contrast, if $\bm{t}_k$ corresponds to the other (interference) sources, $\bm{z}_k$ should be a one-hot vector.
This allocation can be interpreted as a partitional clustering of the basis vectors.
Although such optimization can be facilitated by~\eqref{eq:propNtf} and the low-rank approximation property in NTF, we introduce a new attractor-based regularization to enhance the aforementioned partitional clustering further.

The optimization problem of the proposed method is formulated as follows:
\begin{align}
    \underset{\bm{Z}, \bm{T}, \bm{V}}{\mathrm{minimize}}\!\!~& \tsum_{a,i,j}\!\mathcal{D}\!\left(\!c^\mathrm{(prop)}_{a,i,j} | \tsum_k\! z_{a,k}t_{i,k}v_{j,k}\! \right) \!+\! \mu \!\tsum_{k}\!\mathcal{R}\!\left(\bm{p}_{b_k} | \bm{z}_{k} \right) \nonumber \\
    &~~~\mathrm{s.t.}~z_{a,k}, t_{i,k}, v_{i,k}, \geq 0~\forall a,i,j,k, \label{eq:propCost}
\end{align}
where $\mu \geq 0$ is the weight coefficient, and the regularization term is defined as 
\begin{align}
    \mathcal{R}\left(\bm{p}_{b_k} | \bm{z}_{k} \right) &= \tsum_{a} \mathcal{D}(p_{a, b_k} | z_{a, k}), \label{eq:propReg} \\
    \mathbb{P} &= \left\{
        \bm{p}_0, \bm{p}_1, \cdots, \bm{p}_{B-1}
    \right\}, \\ 
        \bm{p}_0 &\coloneqq
        [
        1 / A, 
        1 / A, 
        \cdots, 
        1 / A
        ]^\mathrm{T}\in\{1/A\}^A, \nonumber \\
    \bm{p}_1 &\coloneqq
        [
        1, 
        0, 
        \cdots, 
        0 
        ]^\mathrm{T}\in\{0, 1\}^A, \nonumber \\
    \bm{p}_2 &\coloneqq 
        [
        0, 
        1, 
        \cdots, 
        0 
        ]^\mathrm{T}\in\{0, 1\}^A, \nonumber \\
        &~~\! \vdots \nonumber \\
    \bm{p}_{B-1} &\coloneqq
        [
        0, 
        0, 
        \cdots,
        1 
        ]^\mathrm{T}\in\{0, 1\}^A, \nonumber \\
    b_k &\in \underset{b}{\mathrm{argmin}}~\tsum_{a} \mathcal{D}(p_{a,b}|z_{a,k}). \label{eq:bk}
\end{align}
Also, $p_{a, b}$ is an element of $\bm{p}_b$, $b=0, 1,\cdots, B-1$ is the index of attractor vectors $(\bm{p}_0, \bm{p}_1, \cdots, \bm{p}_{B-1})$, and $B=A+1$.
The set $\mathbb{P}$ encompasses $B$ attractor vectors for each class; $\bm{p}_{0}$ corresponds to the target-source class, while $\bm{p}_{1}, \bm{p}_2, \cdots, \bm{p}_{B-1}$ correspond to the other interference-source classes related to each microphone array.
$b_k$ calculated by~\eqref{eq:bk} corresponds to the index of the nearest attractor vector (class) from the current allocation vector $\bm{z}_k$. 
Thus, the regularization term in~\eqref{eq:propReg} forces $\bm{z}_k$ to be closer to the nearest attractor vector, $\bm{p}_{b_k}$, emphasizing source clustering of the basis vectors $(\bm{t}_k)_{k=0}^{K-1}$. 
This can further affect the basis matrix, resulting in more discriminative basis vectors. 
In addition, this regularization automatically classifies $K$ basis vectors into target and interference sources.
Consequently, the optimal number of basis vectors for the target source is estimated jointly during the optimization.

After the optimization of $\bm{Z}$, $\bm{T}$, and $\bm{V}$, a binary vector $\bm{h}=[h_0, h_1, \cdots, h_{K-1}]^\mathrm{T}\in\{0, 1\}^K$ is calculated as follows:
\begin{align}
    h_k =
    \begin{cases}
        1 & ~(\mathrm{if}~b_k=0) \\
        0 & ~(\mathrm{otherwise})
    \end{cases}, \label{eq:binaryVec}
\end{align}
where $h_k=1$ indicates that $\bm{t}_k$ corresponds to the target source.
In contrast to~\eqref{eq:convMask}, \eqref{eq:binaryVec} is independent of the time frame $j$.

Similar to~\eqref{eq:convWiener}, the spectrogram of the target source is obtained using the following Wiener filter: 
\begin{align}
    \hat{s}^{(a)}_{i,j} = \frac{\tsum_k{(h_{k}z_{a,k}t_{i,k}v_{j,k})^2}}{\tsum_k({z_{a,k}t_{i,k}v_{j,k}})^2}y^{(a)}_{i,j}. \label{eq:propWiener}
\end{align}
The other post-processing steps are the same as those used in the conventional method.

\subsection{Derivation of Update Rules}

The cost function in~\eqref{eq:propCost} can be minimized using a majorization-minimization (MM) algorithm~\cite{Hunter2000_mmAlg}.
In this study, we use a generalized Kullback--Leibler divergence 
\begin{align}
    \mathcal{D}(\mathrm{b}|\mathrm{a}) = \mathrm{b} \log{\frac{\mathrm{b}}{\mathrm{a}}} + \mathrm{a} - \mathrm{b}
\end{align}
in~\eqref{eq:convCost}, \eqref{eq:propCost}, and \eqref{eq:bk}, which provides better performance in many audio source separation tasks, e.g.,~\cite{Kitamura2014_snmf}. 
We define the cost function in~\eqref{eq:propCost} as $\mathcal{J}$.
By applying Jensen's inequality, we obtain the majorization function $\mathcal{J}^+ \geq \mathcal{J}$ as follows:
\begin{align}
    \mathcal{J}^+ &\overset{c}{=} \tsum_{a,i,j} \left(-c^\mathrm{(prop)}_{a,i,j} \tsum_k \alpha_{a,i,j,k} \log{ \frac{z_{a,k} t_{i,k} v_{j,k}}{\alpha_{a,i,j,k}}} \right. \nonumber \\
    &\phantom{=} \mbox{} \left.+ \tsum_k z_{a,k}t_{i,k}v_{j,k}\right) + \mu\tsum_{a,k}\left(-p_{a, b_k}\log z_{a,k} + z_{a,k}\right),
\end{align}
where $\overset{c}{=}$ denotes equality up to a constant and $\alpha_{a,i,j,k} > 0$ is an auxiliary variable that satisfies $\sum_k \alpha_{a,i,j,k} = 1$.
The equality $\mathcal{J}^+ = \mathcal{J}$ holds if and only if 
\begin{align}
    \alpha_{a,i,j,k} = \frac{ z_{a,k}t_{i,k}v_{j,k}}{ \sum_{k'} z_{a, k'}t_{i, k'}z_{j, k'} }. \label{eq:propAlpha}
\end{align}
From $\partial \mathcal{J}^+ / \partial z_{a, k} = 0$, we obtain
\begin{align}
    \tsum_{i,j}\left(-c^\mathrm{(prop)}_{a,i,j} \frac{\alpha_{a,i,j,k}}{z_{a,k}} + t_{i,k}v_{j,k} \right) + \mu\left(-\frac{p_{a,b_k}}{z_{a,k}} + 1\right)= 0. \nonumber
\end{align}
Thus, 
\begin{align}
    z_{a,k} = \frac{ \tsum_{i,j} c^\mathrm{(prop)}_{a,i,j} \alpha_{a,i,j,k} + \mu p_{a,b_k} }{ \tsum_{i,j}t_{i,k}v_{j,k} + \mu }. \label{eq:propZderivative}
\end{align}
By substituting~\eqref{eq:propAlpha} into~\eqref{eq:propZderivative}, we obtain the update rule for $\bm{Z}$ as
\begin{align}
    z_{a,k} \leftarrow \frac{ z_{a,k} \tsum_{i,j} c^\mathrm{(prop)}_{a,i,j} \frac{t_{i,k}v_{j,k}}{\tsum_{k'} z_{a,k'}t_{i,k'}v_{j,k'}} + \mu p_{a,b_k} }{\tsum_{i,j} t_{i,k}v_{j,k} + \mu}. \label{eq:updZ}
\end{align}
Note that index $b_k$ must always be updated by~\eqref{eq:bk} before updating $\bm{Z}$.
Similar to~\eqref{eq:updZ}, the update rules for $\bm{T}$ and $\bm{V}$ can be derived as follows:
\begin{align}
    t_{i,k} &\leftarrow t_{i,k} \frac{ \tsum_{a,j} c^\mathrm{(prop)}_{a,i,j} \frac{z_{a,k}v_{j,k}}{\tsum_{k'}z_{a,k'}t_{i,k'}v_{j,k'}}}{\tsum_{a,j} z_{a,k}v_{j,k}}, \label{eq:updT} \\
    v_{j,k} &\leftarrow v_{j,k} \frac{ \tsum_{a,i} c^\mathrm{(prop)}_{a,i,j} \frac{z_{a,k}t_{i,k}}{\tsum_{k'} z_{a,k'}t_{i,k'}v_{j,k'}}}{\tsum_{a,i} z_{a,k}t_{i,k}}. \label{eq:updV}
\end{align}
To ensure $\tsum_a z_{a,k}=1$ and $\tsum_i t_{i,k} = 1$, we apply the normalization of $\bm{z}_{k}$ and $\bm{t}_k$ after \eqref{eq:updZ} and \eqref{eq:updT}, respectively, such that the cost function does not change by scaling each column of $\bm{V}$.

For the convergence of the proposed optimization algorithm, we can state the following theorem. 
This ensures a theoretical nonincrease in the cost function in the proposed method.
\begin{thm}
The update rules~\eqref{eq:bk}, \eqref{eq:updZ}--\eqref{eq:updV} ensure the monotonic nonincrease of the cost function in \eqref{eq:propCost}.
\end{thm}
\begin{proof}
Based on the MM algorithm, update rules \eqref{eq:updZ}--\eqref{eq:updV} ensure a monotonic nonincrease of the cost function in \eqref{eq:propCost}. 
Thus, the monotonic nonincrease of the entire algorithm \eqref{eq:bk}, \eqref{eq:updZ}--\eqref{eq:updV} depends on whether~\eqref{eq:bk} has a monotonic nonincrease property. 
Let $b_k^{(\mathrm{old})}$ and $b_k^{(\mathrm{new})}$ represent the old and updated indices, respectively.
As $b_k$ is updated such that $\sum_a \mathcal{D}(p_{a,b_k}|z_{a,k})$ is minimized in~\eqref{eq:bk}, the following inequality holds:
\begin{align}
    \mathcal{R} (\bm{p}_{b_k^{(\mathrm{old})}} | \bm{z}_k) &= \sum_{a} \mathcal{D}(p_{a, b_k^{(\mathrm{old})}} | z_{a,k}) \nonumber \\
    &\geq \sum_{a} \mathcal{D}(p_{a, b_k^{(\mathrm{new})}} | z_{a,k}) \nonumber \\
    &= \mathcal{R} (\bm{p}_{b_k^{(\mathrm{new})}} | \bm{z}_k)~~\forall k.
\end{align}
Therefore, the update rule~\eqref{eq:bk} does not increase the value of the cost function in~\eqref{eq:propCost}.
\end{proof}

\section{Experiment}
\label{sect:exp}
\subsection{Conditions}

\begin{figure}[t]
    \begin{center}
        \subfigure[]{\includegraphics[width=0.48\columnwidth]{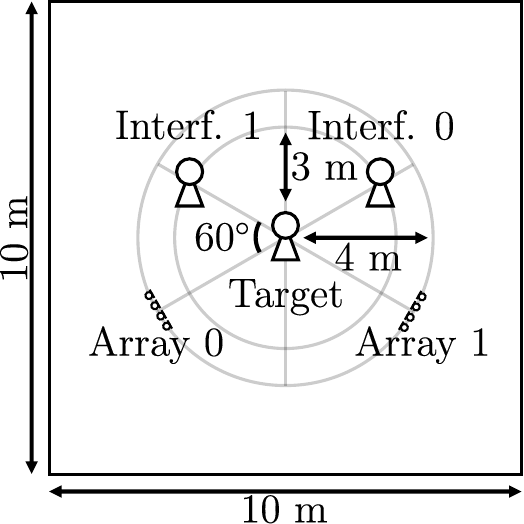}}
        \subfigure[]{\includegraphics[width=0.48\columnwidth]{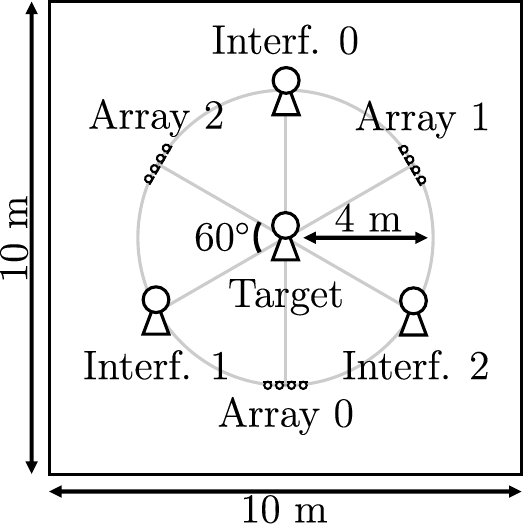}}
    \end{center}
    \vspace{-10pt}
    \caption{Recording environments simulated by the two-dimensional image method: (a) two-microphone-array and (b) three-microphone-array cases. All the microphone spacing in each array is set to 2.83~cm.}
    \label{fig:room}
    \vspace{-5pt}
\end{figure}

A spotforming experiment was conducted to validate the proposed method.
To simulate the recording environment illustrated in Fig.~\ref{fig:room}, we used a two-dimensional image method implemented in Pyroomacoustics~\cite{Scheibler2018_pyroomacoust}.
We simulated two reverberation times for each environment, $T_{60}=0$ and $T_{60}=256$~ms, resulting in four recording conditions. 
The speech signals listed in Table~\ref{tbl:drySoruce}, randomly selected from the LibriTTS~\cite{Zen2019_libritts} dataset, were used as the dry sources. 
These dry sources were normalized to have uniform signal energies before the room impulse responses were convoluted.
The effect of background noise was not considered in this experiment.

We applied a minimum variance distortionless response (MVDR) BF to each observed signal $\mathsf{X}^{(a)}$ of the microphone array and obtained enhanced signals $(\bm{Y}^{(a)})_{a=0}^{A-1}$. 
The target steering vectors and noise covariance matrices for each MVDR BF were set to their oracle values calculated from impulse responses.
This condition simulated that BF preprocessing provides ideal performance, and the net performances of the conventional and proposed methods were compared. 
All microphones were synchronized in this experiment.

As an evaluation criterion, we used the source-to-distortion ratio (SDR)~\cite{Vincent2006_bsseval} of the target source, a common score reflecting the total source separation quality. 
Because the NMF and NTF results depend on the initial random values of the parameters, we used 10 random seeds. 
We calculated the average SDR scores and their standard deviations.
The other conditions are listed in Table~\ref{tbl:expCond}.

\subsection{Results and Discussion}

\begin{figure*}[t]
    \vspace{-5pt}
    \begin{center}
        \subfigure[]{\includegraphics[width=0.49\textwidth]{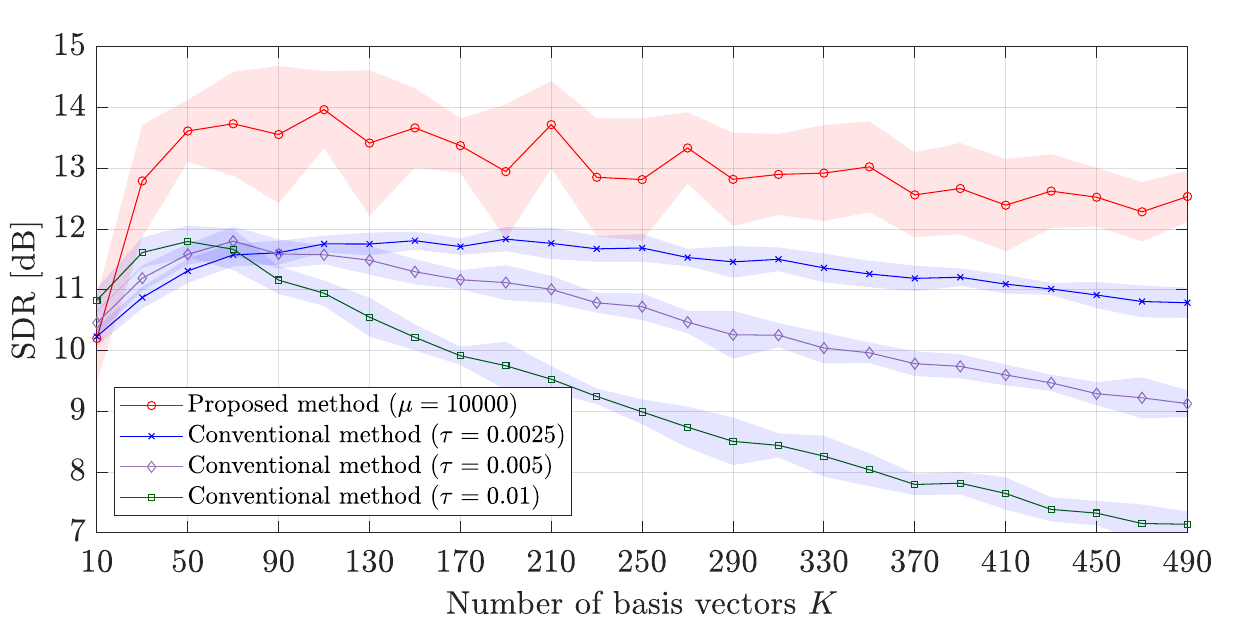}}
        \subfigure[]{\includegraphics[width=0.49\textwidth]{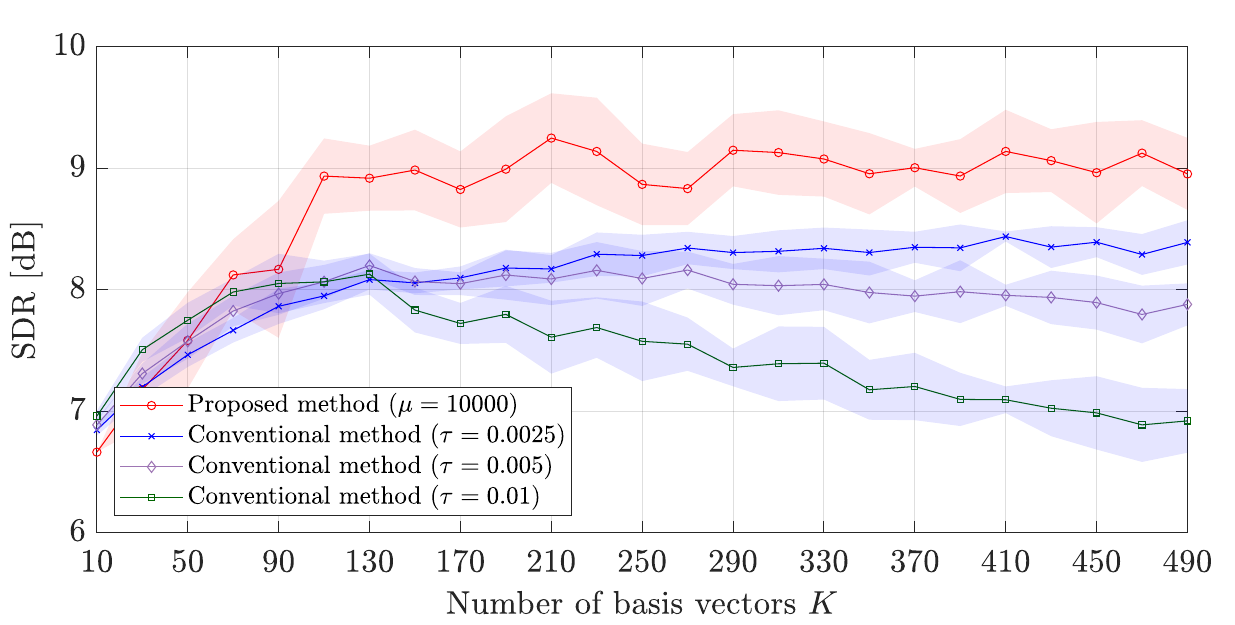}}
    \end{center}
    \vspace{-10pt}
    \caption{SDR scores with various $K$ in the two-microphone-array case: (a) $T_{60}=0$~ms and (b) $T_{60}=256$~ms. The plots and colored areas show the average values and standard deviations. Average SDRs of simple BF outputs $\bm{Y}^{(0)}$ and $\bm{Y}^{(1)}$ were 9.4~dB in (a) and 6.7~dB in (b).}
    \label{fig:exp2arrays}
\end{figure*}

\begin{figure*}[t]
    \vspace{-5pt}
    \begin{center}
        \subfigure[]{\includegraphics[width=0.49\textwidth]{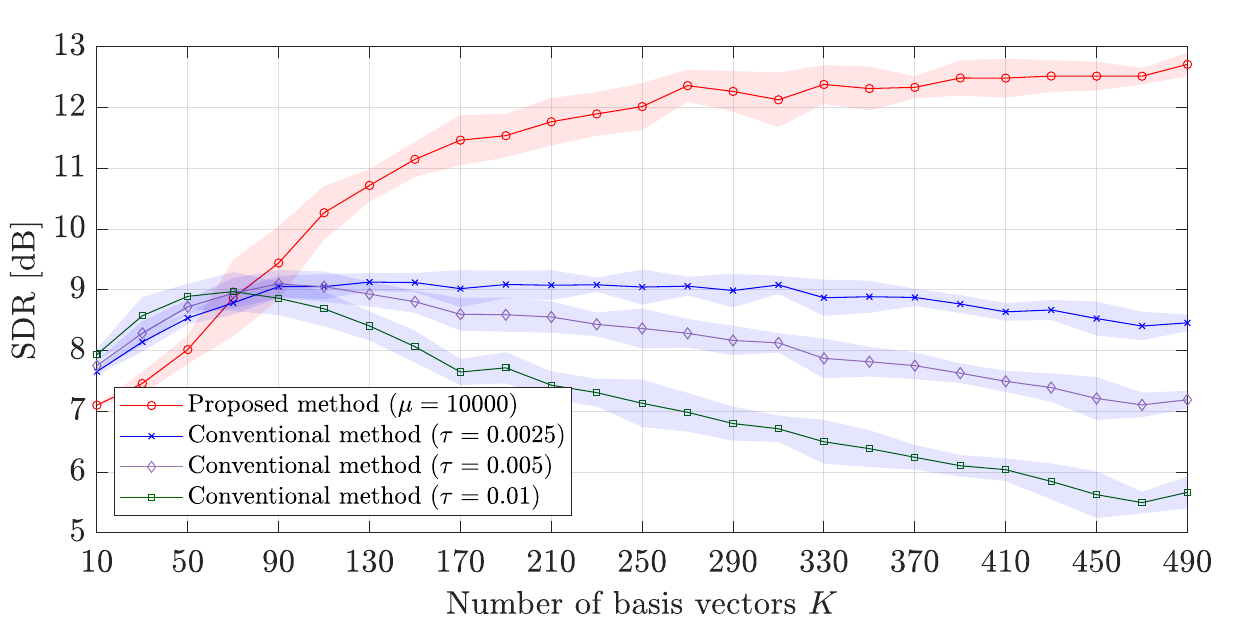}}
        \subfigure[]{\includegraphics[width=0.49\textwidth]{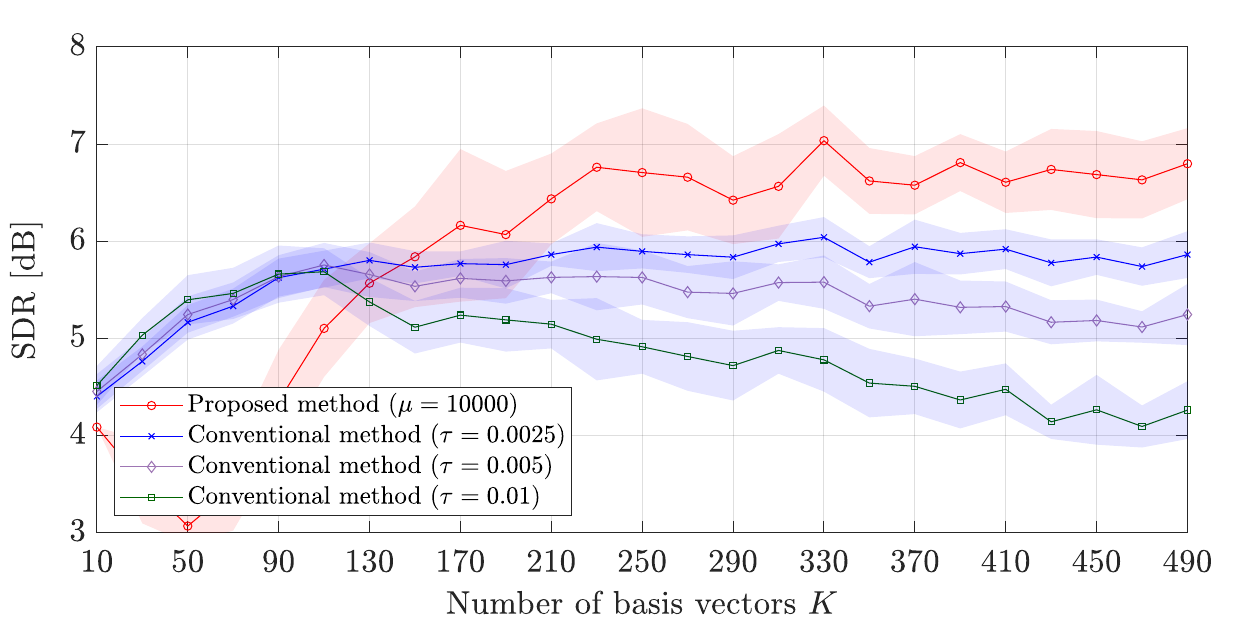}}
    \end{center}
    \vspace{-10pt}
    \caption{SDR scores with various $K$ in the three-microphone-array case: (a) $T_{60}=0$~ms and (b) $T_{60}=256$~ms. The plots and colored areas show the average values and standard deviations. Average SDRs of simple BF outputs $\bm{Y}^{(0)}$, $\bm{Y}^{(1)}$, and $\bm{Y}^{(2)}$ were 7.1~dB in (a) and 4.1~dB in (b).}
    \label{fig:exp3arrays}
\end{figure*}

Figs.~\ref{fig:exp2arrays} and \ref{fig:exp3arrays} show the results of two- and three-microphone-array cases, respectively. 
The hyperparameter $\tau$ for the conventional method was set to 12 patterns, and the three better conditions were shown in Figs.~\ref{fig:exp2arrays} and \ref{fig:exp3arrays}.
These results confirm that the proposed method outperforms the conventional method in all cases.
Moreover, the proposed method maintained better performance when $K$ increased, while the performance of the conventional method degraded for a large value of $K$ in some conditions of $\tau$.

The hyperparameter $\mu$ in the proposed method also affected the performance. 
Fig.~\ref{fig:expMu} shows the performance behavior of the proposed method with various settings of $\mu$.
The proposed method achieved optimal performance when we set $\mu$ to a certain large value, e.g., $\mu \geq 100$.
Strong regularization with~\eqref{eq:propReg} results in a hard classification of the basis vectors $(\bm{t}_k)_{k=0}^{K-1}$, namely, each allocation vector $\bm{z}_k$ coincides with one of the attractor vectors $(\bm{p}_b)_{b=0}^{B-1}$.
This phenomenon was consistently confirmed under other values of $K$ and $T_{60}$.
Thus, such optimization tends to provide a better spotforming performance for the proposed method.
Thanks to this property, we can robustly obtain better results by using a certain large value of $\mu$ and $K$.

\begin{table}[t]
    \vspace{-5pt}
    \caption{Dry sources}
    \label{tbl:drySoruce}
    \begin{center}
        \vspace{-10pt}
        \scalebox{0.8}{
            \begin{tabular}{cc}\Hline
                File name                                   & Source \\ \hline
                \texttt{84\_121123\_000008\_000002.wav}     & Target \\
                \texttt{652\_130737\_000012\_000000.wav}    & Interf.~0 \\
                \texttt{3000\_15664\_000020\_000005.wav}    & Interf.~1 \\
                \texttt{1272\_141231\_000024\_000005.wav}   & Interf.~2 \\ \Hline
            \end{tabular}
        }
        \vspace{0pt}
    \end{center}
    \vspace{-17pt}
\end{table}

\begin{table}[t]
    \vspace{-11pt}
    \caption{Experimental conditions}
    \label{tbl:expCond}
    \begin{center}
        \vspace{-12pt}
        \scalebox{0.8}{
            \begin{tabular}{ll}\Hline
                \raisebox{-0.2ex}[0cm][0cm]{Sampling frequency}                          & \raisebox{-0.2ex}[0cm][0cm]{Down sampled to 16~kHz} \\ \hline
                \raisebox{-0.2ex}[0cm][0cm]{Window function used in STFT}                & \raisebox{-0.2ex}[0cm][0cm]{Hann window} \\ \hline
                \raisebox{-0.2ex}[0cm][0cm]{Window length in STFT}                       & \raisebox{-0.2ex}[0cm][0cm]{32~ms} \\ \hline
                \raisebox{-0.2ex}[0cm][0cm]{Window shift length in STFT}                 & \raisebox{-0.2ex}[0cm][0cm]{16~ms} \\ \hline
                \raisebox{-0.2ex}[0cm][0cm]{Number of iterations in NMF/NTF}             & \raisebox{-0.2ex}[0cm][0cm]{100 times} \\ \hline
                \raisebox{-0.2ex}[2ex][0.1cm]{Initial values of $\bm{T}$, $\tilde{\bm{V}}$, and $\bm{V}$} & \raisebox{-0.0ex}[0cm][0cm]{Uniform random values in the range $(0, 1)$} \\ \hline
                \raisebox{-0.2ex}[0cm][0cm]{Initial values of $\bm{Z}$}                  & \raisebox{-0.2ex}[0cm][0cm]{All the elements are set to $1/A$} \\ \hline
                \multirow{2}{*}{Weight coefficient $\mu$}   & \raisebox{-0.2ex}[0cm][0cm]{$\mu=0$ for first 50 iterations, and}\\ 
                                                            & \raisebox{-0.2ex}[0cm][0cm]{$\mu>0$ for the rest of iterations} \\ \Hline
            \end{tabular}
        }
        \vspace{0pt}
    \end{center}
    \vspace{-16pt}
\end{table}

\section{Conclusion}

This study has proposed a new spotforming algorithm that unifies NTF and attractor-based regularization. 
The regularization term is designed based on the partitional clustering of the NTF basis vectors into target and interference sources.
The experimental results revealed that the proposed method outperformed the conventional NMF-based spotforming technique.

\end{document}